\documentclass{llncs}
\usepackage[english]{babel} 
\usepackage{amsmath,amssymb,amsfonts,mathrsfs}
\usepackage[T1]{fontenc}
\usepackage[utf8]{inputenc}
\usepackage{booktabs}
\usepackage{algorithmic}
\usepackage{algorithm}

\spnewtheorem{assumption}{Assumption}{\bfseries}{\itshape}
\spnewtheorem*{lemma*}{Lemma}{\normalshape\bfseries}{\itshape}

\newcommand{\eqdef}{\triangleq}
\renewcommand{\leq}{\leqslant}
\renewcommand{\geq}{\geqslant}
\renewcommand{\epsilon}{\varepsilon}
\newcommand{\rg}[2]{\left \{ #1,\dots,#2  \right \}}
\newcommand{\affect}{\leftarrow}
\newcommand{\raffect}{\xleftarrow{\$}}
\newcommand{\support}[2]{\langle {#2} {\rangle}_{{#1}}}
\newcommand{\sphere}{\mathbb{S}}
\newcommand{\grassman}[3]{\boldsymbol{\mathrm{Gr}}_{#1}(#2,#3)}

\DeclareMathOperator{\prob}{\mathbb{P}}
\DeclareMathOperator{\dec}{\Phi}

\newcommand{\F}{\mathbb{F}}
\newcommand{\fq}{\F_q}
\newcommand{\fqm}{\F_{q^m}}
\newcommand{\set}[1]{\mathcal{#1}}
\newcommand{\card}[1]{\left \lvert  {#1} \right \rvert }
\newcommand{\bigcard}[1]{\Big \lvert  {#1} \Big \rvert }
\newcommand{\MS}[3]{#3^{#1 \times #2}}%matrix set

\newcommand{\word}[1]{\boldsymbol{\mathrm{#1}}}
\newcommand{\av}{\word{a}}
\newcommand{\bv}{\word{b}}
\newcommand{\ev}{\word{e}}
\newcommand{\hv}{\word{h}}
\newcommand{\sv}{\word{s}}
\newcommand{\uv}{\word{u}}
\newcommand{\xv}{\word{x}}

\newcommand{\mat}[1]{\boldsymbol{\mathrm{#1}}}
\newcommand{\Am}{\mat{A}}
\newcommand{\Bm}{\mat{B}}
\newcommand{\Em}{\mat{E}}
\newcommand{\Hm}{\mat{H}}
\newcommand{\Mm}{\mat{M}}
\newcommand{\Rm}{\mat{R}}
\newcommand{\Sm}{\mat{S}}
\newcommand{\Um}{\mat{U}}
\newcommand{\Vm}{\mat{V}}

\begin{document}
                                 
\title{An Upper-Bound on the Decoding Failure Probability of the LRPC Decoder}	
\author{\'Etienne Burle  \and Ayoub Otmani}
\institute{LITIS, University of Rouen Normandie, France
\email{\{Etienne.Burle,Ayoub.Otmani\}@univ-rouen.fr}}
	
\maketitle	

\begin{abstract}
Low Rank Parity Check (LRPC) codes form a class of rank-metric error-correcting codes that was purposely introduced to design  public-key encryption schemes. An LRPC code is defined from a parity check matrix  whose entries belong to a relatively low dimensional vector subspace of a large finite field. This particular algebraic feature can then be exploited  to correct with high probability rank errors  when the   parameters are  appropriately chosen. In this paper,  we present theoretical upper-bounds on the  probability that the LRPC decoding algorithm fails.
	
\end{abstract}

\section{Introduction}

Rank-metric cryptography has attracted a relative interest over the last years mainly  thanks to the recent trend that appeared with the goal of standardizing quantum-safe public-key algorithms. ROLLO \cite{ABDGHRTZABBBO19} and RQC \cite{AABBBDGZCH19} are  two examples of rank-metric public-key encryption schemes that were submitted to the NIST call for standardizing  quantum-resistant public-key cryptographic algorithms.  
The theory of codes endowed with the rank-metric was first studied in \cite{d78} where a Singleton-type bound was proved and a class of codes reaching the bound was given. A few years later, Gabidulin constructed   \cite{G85} the first example of rank-metric error-correcting codes which can be seen as the counterparts of generalized Reed-Solomon (GRS) codes. The so-called Gabidulin codes are defined from the evaluation of non-commutative linearized polynomials \cite{O33}. 
They  can be efficiently decoded by an equivalent of the Euclidean algorithm \cite{O33a} while achieving the rank-Singleton upper-bound.
Not long after, the first rank-metric public-key encryption scheme called the GPT cryptosystem appeared  in \cite{GPT91}.
It bore strong similarities with the famous McEliece cryptosystem \cite{M78}.
The GPT scheme is indeed an analogue of the McEliece cryptosystem but based on  Gabidulin codes.  
Not surprisingly, this strong resemblance to GRS codes is the reason why their use in the GPT cryptosystem has been  subject to
several attacks \cite{G95,G96}, as well as the different reparations  that were subsequently cryptanalysed \cite{O05,O05a,O08}.
These flaws in the design do not mean that the rank-metric is not viable in cryptography.
Indeed, the famous decoding problem has naturally its rank version which is also believed to be intractable both in a classical and quantum setting.

ROLLO  replaces Gabidulin codes with the class of (Ideal) Low Rank Parity Check (LRPC) codes introduced in \cite{AGHRZ19a}. 
An LRPC code is defined by means of an homogeneous $(n-k) \times n$ parity-check matrix $\Hm = \begin{bmatrix}h_{i,j}\end{bmatrix}$ where each entry $h_{i,j}$ lies in a linear subspace $\set{W} \subsetneq \fqm$ over $\fq$ of relatively low dimension $w$.
This property can then be exploited to design a probabilistic decoding algorithm that can recover any error vector $\ev \in \fqm^n$ of rank weight $t \leq (n-k)/w$.
 
The principle behind the LRPC decoder \cite{AGHRZ19a} is to view the syndrome  $\sv = \ev \Hm^{\mathsf{T}}$ as a sample of a  uniformly distributed random variable taking values on $\left ( \set{E} \cdot \set{W} \right)^{n-k}$ where $\set{E} \subsetneq \fqm$ is the $t$-dimensional linear space generated over $\fq$ by the coordinates of $\ev$.
Under the assumption that the linear space over $\fq$  spanned by the entries of $\sv$ denoted by  $\set{S}\subset \fqm$ is equal to $\set{E} \cdot \set{W}$, the  decoding algorithm  first recovers a basis $\epsilon_1,\dots,\epsilon_t$ of $\set{E}$ by computing the intersection $\bigcap_{i=1}^w  f_i^{-1} \cdot \set{S}$ where $\{f_1,\cdots,f_w\}$ is an arbitrary (known) basis of $\set{W}$. The success of this step  lies in the fact that with high probability this intersection is equal to  $\set{E}$.
The last step then consists in computing the coordinates $e_1,\dots,e_n$ of $\ev$ by writing that $e_j = \sum_{d=1}^t x_{j,d} \epsilon_d$ where each $x_{j,d} \in \fq$ is unknown. One can then solve the linear system $\sv = \ev \Hm^{\mathsf{T}}$ and expect to find a unique solution when $w \geq n/(n-k)$ because in that case the number of  unknowns  $n t$ is at most the number   $(n-k) wt$ of linear equations.  

Recently, an encryption scheme based on LRPC codes has been proposed in \cite{AADGZ22} where the decoder receives a matrix of syndromes $\Sm = \Em \Hm^{\mathsf{T}}$ where $\Em$ is an homogeneous matrix so that  the probability that the entries of $\Sm$ span  $ \set{E} \cdot \set{W}$ is increased.
Another work \cite{BGHO22}  gives a new construction of error-correcting codes that can be decoded by the same techniques but relies on a generalization of the notion of homogeneous matrices. It introduced the concept of \emph{semi-homogenous} parity-check matrices which are matrices such that the coordinates of each row span a different low-dimensional linear subspace of $\fqm$. 
This enables the authors to build a public-key encryption scheme where the public key is statistically  close to a random matrix. Note that the security of ROLLO relies on the difficulty of the (Ideal) LRPC code \emph{indistinguishability problem} which asserts that it is computationally hard to distinguish  
a randomly drawn parity-check matrix of an Ideal LRPC code from a random  parity-check matrix of an Ideal code. 

All these schemes have to deal with the decryption failures that inherently come from the LRPC decoding algorithm. As an adversary could shatter the security of these schemes if he manages to exploit decryption failures, it is therefore of paramount importance to  lower the decoding failure probability below the desired security threshold. The best existing bounds on the decoding failure probability are given in \cite{AGHRZ19a,ABDGHRTZABBBO19}.
It is stated in \cite{AGHRZ19a} that the decoding failure probability behaves essentially as $q^{-(n-k) +tw}$ which comes from an approximation of the probability that  the entries of the syndrome vector $\sv$ does not span $ \set{E} \cdot \set{W}$. Another analysis is given in \cite{ABDGHRTZABBBO19} resulting to the expression 
$q^{-(n-k)+tw-1} + q^{-(w-1)(m-tw-t)}$. The first term corresponds to a tighter approximation of the one given in \cite{AGHRZ19a}, and the quantity $q^{-(w-1)(m-tw-t)}$ reflects the probability that the intersection of random linear subspaces $\set{R}_1,\dots,\set{R}_w$  all containing $\set{E}$ is different from $\set{E}$. 
Several works \cite{AGHRZ19a,ABDGHRTZABBBO19,AADGZ22} assumed that $f_i^{-1} \cdot \set{S}$ behaves as a random linear space $\set{R}_i $ containing $\set{E}$.
But this hypothesis  cannot be realistic because of the existence of the elements $f_1,\dots,f_w$ in $\fqm$ such that $f_i \cdot \set{R}_i = f_j \cdot \set{R}_j$ for every $i \neq j$ when  $\set{R}_i = f_i^{-1} \cdot \set{S}$. 
Although the validity of the approximation $q^{-(w-1)(m-tw-t)}$ is verified by simulations in \cite{ABDGHRTZABBBO19}, it does not necessarily predict the asymptotic behavior. 
 
\subsection*{Our Contribution and Main Results} 

We revisit the analysis of the LRPC decoder with the main goal to establish provable theoretical  bounds. 
Although we do not reach the best existing heuristic approximations, our  work manages to close a little bit further the gap between the  theoretical  bounds and the  practical approximations.
We provide in  Table \ref{tb:comp} a comparison between existing bounds and the bounds we obtain in this work.

As we have seen, there are several reasons that make the LRPC decoder fail. 
The first one comes from the fact that the entries of $\sv$ might not span $\set{E} \cdot \set{W}$.  In \cite[Proposition 4.3]{AGHRZ19a}, the authors state that the coordinates of $\sv$ are independently and uniformly distributed over  $ \set{E} \cdot \set{W}$ leading them to  upper-bound the probability\footnote{We can also get this result by using directly Theorem~2 from \cite{AADGZ22}.} by $q^{-(n-k)+tw}$. We provide in Proposition \ref{prop:randomUV} a simple argument that explains why the coordinates of $\sv$ are independent and uniform random variables over the random choices over $\Hm$ and $\ev$.
This enables us to use  the closed-form expression of the probability that random vectors belonging to the same linear subspace span it.
We apply this result to the coordinates of the syndrome vector $\sv$ and we show in Proposition \ref{prop:simplified_prod_bound} that this probability is lower than $q^{-(n-k)+tw}/(q-1)$.
We notice that when $\dim(\set{E} \cdot \set{W}) = tw$ the probability is equivalent to this term (see Remark \ref{rk:equiv}). As a consequence, the upper-bound $q^{-(n-k)+tw-1}$ given in \cite{ABDGHRTZABBBO19} cannot hold.	

Next, the second reason why the LRPC decoder might not decode correctly comes from the fact that we do not obtain $\set{E}$ when computing  $\bigcap_{i=1}^w  f_i^{-1} \cdot \set{S}$. 
In the literature there exists essentially two ways to upper-bound the probability of occurrence of this event. 
One approach is described in \cite{AGHRZ19a} where two upper-bounds are given: in \cite[Proposition 3.5]{AGHRZ19a} the probability is at most $t q^{tw(w+1)/2-m}$ and in \cite[Proposition 3.8]{AGHRZ19a} it is at most $t q^{(2w-1)t-m}$.
The other path followed in \cite[Proposition 2.4.2]{ABDGHRTZABBBO19} and \cite[Proposition 3]{AADGZ22}  consists in assuming as explained previously that $f_i^{-1} \cdot \set{S}$ behaves as a random linear space $\set{R}_i $ containing $\set{E}$.
This enables the authors  to prove that the probability   is at most $q^{-(w-1)(m-tw-t)}$.
In this work, we depart from this assumption and we prove in Theorem \ref{th:bound:intersection}  that this probability is at most $q^{(2w   - 1)t}/(q^m - q^{t-1})$. Although our bound is less interesting than  $q^{-(w-1)(m-tw-t)}$, it is however better than the theoretical ones given in \cite{AGHRZ19a}.

Finally the last situation that induces a decoding failure is when the unknown coordinates of $\ev$ cannot be recovered because the linear system inferred from $\sv = \ev \Hm^{\mathsf{T}}$  is not of full rank. This  happens when the dimension of $\set{E} \cdot \set{W}$ is strictly less than  $ \dim_{\fq} \set{E} \dim_{\fq} \set{W} = tw$. 
The paper \cite{AGHRZ19a} shows in Proposition 3.3 that this case happens with  probability at most $t q^{tw -m}$ over the random choice of $\set{E}$ and for a given set $\set{W}$. 
In Proposition \ref{prop:dimEW = wt} we improve this bound by showing that this probability is at most $q^{tw}/\left(q^m - q^{t-1}\right)$.

Theorem \ref{th:decoding}  summarizes all our theoretical analysis which allows us to prove that when $twq^{-(n-k)+tw}\leq 1$, $tw = \omega(1)$ and $k = \Theta(n)$, we obtain an upper-bound asymptotically equivalent to $q^{-(n-k)+tw}/(q-1)+ q^{2tw - m}$ as $n$ tends to $+\infty$ (Corollary \ref{cor:asymp}).

\begin{table}
	\begin{center}	
		\begin{tabular}{*{5}{l}} \toprule
			Case of error & &     Previous bound \cite{AGHRZ19a} & \hspace*{0.5cm} & Our bound \\ \midrule
			$\prob \left \{ \support{\fq}{\ev\Hm^\mathsf{T}}  \neq \set{E} \cdot \set{W} \right\}$  &  & $q^{-(n-k) +tw}$ &  & $1 -  \prod\limits_{i=0}^{tw-1} \left(1 - q^{i-(n-k)} \right)<\frac{q^{-(n-k)+tw}}{q-1}$ \\ 
			\midrule 
			$\prob 
			\left \{ 
			\set{E} \neq   \bigcap_{i=1}^w  f_i^{-1}  \cdot \set{W} \cdot \set{E} 
			\right\}$ & & $t q^{(2w-1)t-m}$ & & $q^{(2w   - 1)t}/(q^m - q^{t-1})$ \\ 
			\midrule
			$\prob\Big\{ \dim \; \set{E} \cdot \set{W} \ne tw \Big\}$ & & $t q^{tw -m}$&  & $q^{tw}/\left(q^m - q^{t-1}\right)$ \\  
			\bottomrule
		\end{tabular}
	\end{center}
	\caption{Comparison with previous theoretical bounds}\label{tb:comp}
\end{table}

\section{Preliminaries}

\subsection{Notation}

The symbol $\eqdef$ will be used to define the left-hand side object. $\card{\set{S}}$ denotes  the cardinality of a set $\set{S}$.
We shall write $x \raffect \set{S}$ to express that $x$ is sampled according to the uniform distribution over a set $\set{S}$.  
We will use the notation $\prob \{ E(x)  \; \vert \; x \raffect \set{S}  \}$ to give the probability that an event $E(x)$ occurs under the constraint that $x \raffect \set{S} $. 
The finite field with $q$ elements where $q$ is a power of a prime number is written as $\fq$.
All vectors will be regarded by default as row vectors and  denoted by boldface letters like $\av = (a_1,\dots{},a_n)$. 
The linear space over a field $\F$  spanned by vectors $\bv_1,\dots,\bv_k$ is written as $\support{\F}{\bv_1,\dots,\bv_k}$.
For  $f \in \F$ and $\set{U} \subseteq \F$, the set $\{ f u \mid u \in \set{U} \}$ is denoted by $f \cdot \set{U} $.
Given two arbitrary sets $\set{A}$, $\set{B}$ included in $\fqm$ where $m \geq 1$, we let 
$\set{A} \cdot \set{B} \eqdef \support{\fq}{a b \mid a \in \set{A}, b \in \set{B}}$.
The set of $r \times n$ matrices with entries in a set $\set{V} \subseteq \F$ is denoted by $\MS{r}{n}{\set{V}}$. 
The transpose is denoted by $^{\mathsf{T}}$. 
For matrices $\Am$ and $\Bm$ having the same number of rows, $\begin{bmatrix}\Am & \vert & \Bm \end{bmatrix}$ represents the matrix obtained by concatenating the columns of $\Am$ followed by the columns of $\Bm$. 

\subsection{Rank Metric}
We consider a finite field extension $\fqm/\fq$ of degree $m \geq 1$ where $q$ is a power of a prime number.  
The \emph{support} of a vector $\xv \in \fqm^L$ denoted by $\support{\fq}{\xv}$ is  the vector space over $\fq$ spanned by its entries, namely
\[
\support{\fq}{\xv} \eqdef \support{\fq}{x_1,\dots{},x_L} \subseteq \fqm.
\]  
The \emph{rank weight} of $\xv$ is then $\dim \support{\fq}{\xv}$.
We let $\grassman{t}{q}{m}$ be the set of all $t$-dimensional linear subspaces over $\fq$ included in $\fqm$. The cardinality of $\grassman{t}{q}{m}$ is given by the Gaussian coefficient: 
\begin{equation}
	\bigcard{\grassman{t}{q}{m}} = \prod_{i=0}^{t-1} \frac{q^{m} - q^i}{q^{t} - q^i}.
\end{equation}
The \emph{sphere} in $\fqm^L$ of radius $w$ centered at $\word{0}$ is denoted by $\sphere_{t}\left(\fqm^L\right)$.
Notice that if $ (\beta_1,\dots,\beta_t)$ is a basis of  $\set{E} \eqdef \support{\fq}{\xv}$ where $\xv \in \sphere_{t}\left(\fqm^L\right)$ then there exists $\Mm \in \MS{t}{L}{\fq}$ such that $\xv = (\beta_1,\dots,\beta_t) \Mm$. 

\subsection{Auxiliary Results}

We gather in this part some results that will be useful  in the next sections.

\begin{proposition} \label{prop:randomUV}
	Let $N$, $L$, $r$ be natural numbers, and consider two independent and uniformly distributed random matrices $\Um \raffect \MS{N}{(L+r)}{\fq}$ and $\Vm \raffect \MS{(L+r)}{r}{\fq}$ with the assumption that $\Vm$ has rank $r$. Then the entries of $\Um \Vm$ are independent and uniformly distributed random variables.
\end{proposition}

\begin{proof}
	Let us write $\Um \Vm = \Um_1 \Vm_1 + \Um_2 \Vm_2$ where
	$\Um = \begin{bmatrix}\Um_1 & \vert & \Um_2 \end{bmatrix}$ with $\Um_1 \in \MS{N}{r}{\fq}$, 
	$\Um_2 \in \MS{N}{L}{\fq}$, and $\Vm = \begin{bmatrix}\Vm_1 \\ \Vm_2 \end{bmatrix}$ with $\Vm_1 \in \MS{r}{r}{\fq}$, 
	$\Vm_2 \in \MS{L}{r}{\fq}$.
	Without loss of generality we can assume that $\Vm_1$ is non-singular and because $\Um_1$ is a uniform random matrix,
	$\Um_1\Vm_1$ is consequently a uniformly distributed random matrix.
	The fact that  $\Um \Vm$ is a uniform random matrix can be inferred from the uniform randomness of $\Um_1\Vm_1$ and the independence between $\Um_2 \Vm_2$ from $\Um_1\Vm_1$.
\qed\end{proof}

\begin{proposition}\label{prop:probUnif}
	Let $\set{U}$ be a vector space of dimension at most $d$  over $\fq$ and  consider an integer $n \geq d$. The probability that $n$ vectors drawn independently and uniformly at random $\uv_1\raffect \set{U},\dots,\uv_{n} \raffect \set{U}$ span $\set{U}$ over $\fq$  is  at least
	\[
	\prod_{i=0}^{d-1} \left(1 - q^{i-n} \right).
	\]
\end{proposition}

\begin{proof}
	Let $a$ be the dimension of $\set{U}$.
	The probability that $\uv_1,\dots,\uv_{n}$ span $\set{U}$ is equal to the probability that an $n\times a$ random matrix with entries in $\fq$ has rank $a$. This probability is then given by
		\[
		\frac{\left(q^n - 1 \right)}{q^n} \times \cdots \times \frac{\left(q^n - q^{a-1} \right)}{q^n}
		\;\; \geq \;\; \frac{\left(q^n - 1 \right)}{q^n} \times \cdots \times \frac{\left(q^n - q^{d-1} \right)}{q^n}
		\]
	where the inequality is derived from the hypothesis that $a \leq d$.
	\qed
\end{proof}

\begin{remark}\label{rk:a=d}
	The inequality in Proposition \ref{prop:probUnif} is an equality if the dimension of $\set{U}$ equals $d$.
\end{remark}

\begin{lemma} \label{lem:aiei=0}
	Consider $\set{A} \subseteq \support{\fq}{g_1,\dots,g_{r}} \subset \fqm$  with $0 < r \leq m$, and  let us assume that $\set{A}$ contains a linear space over $\fq$ of dimension at least $d \leq r$.
	For randomly drawn elements $e_1,\dots,e_t$ from $\fqm$  such that  $\support{\fq}{e_1,\dots{},e_t} \in \grassman{t}{q}{m}$, there exists $\av$ in $\set{A}^t \setminus \{\word{0}\}$ such that $\sum_{j=1}^t a_j e_j = 0$ with probability at most 
	\begin{equation} \label{bound:aiei=0}
		\frac{ q^{tr+1} - \left(q^d -1 \right) \left(q^{t} - 1\right) }{q^{m+1} - q^{t}}
	\end{equation}
\end{lemma}

\begin{proof}
	Let us fix  an arbitrary $t$-tuple $\av = (a_1,\dots,a_t)$ from $\set{A}^t\setminus \{\word{0}\}$. 
	There exists then $i \in \rg{1}{t}$ such that $a_i \neq 0$. The condition $\sum_{j=1}^t a_j e_j = 0$ is equivalent to writing that
	\begin{equation} \label{e_t=sum_iaiei}
		e_i = - a_i^{-1} \sum_{j\neq i} a_j e_j.
	\end{equation}
	Knowing that $e_1,\dots,e_t$ are random elements picked from $\fqm$ that are linearly independent, we can see that we would have a contradiction if $a_j a_i^{-1}$ lies within $\fq$ for every $j$ different from $i$.
	Consequently, we  introduce the set $\set{T} \subsetneq \set{A}^t \setminus \{\word{0}\}$ that \emph{does not} contain $t$-tuples $\av$ such that there exist $i\in \rg{1}{t}$ and scalars $\lambda_j$ from $\fq$ so that we have both $a_i \neq 0$ and $a_j = \lambda_j a_i$ for every $j$ in  $\rg{1}{t}\setminus\{i\}$.
	We can remark that the number of $t$-tuples is least\footnote{Such $t$-tuples $\av$ are of the form $(0,\dots,0,a_i,\lambda_{i+1}a_i,\dots,\lambda_ta_i)$ where $i$ can take any value in $\rg{1}{t}$, $a_i$ is any non-zero element in $\set{A}$, and $\lambda_{i+1},\dots,\lambda_t$ have arbitrary values in $\fq$. The number of such tuples is therefore at least $\left(q^{d} - 1\right) \sum_{u=0}^{t-1} q^u$ because  the choice over $a_i$ can be restricted to the linear space of dimension $d$ contained in the set $\set{A}$.} $\left(q^{d} - 1\right) \left(q^{t} -1\right) / (q - 1)$ and therefore the cardinality of $\set{T}$ is  at most $q^{rt}- 1 - \left(q^{d} - 1\right) \left(q^{t} -1\right) / (q - 1)$.
	From the whole previous discussion, and after applying the Union bound, the probability that we are looking for can be upper-bounded as follows  
	\begin{align*}
		\prob 
		\left \{ 
		\exists \av \in \set{A}^t\setminus \{\word{0}\}, \;\; \sum_{i=1}^t a_i e_i = 0   
		\right\} 
		&=
		\prob 
		\left \{ 
		\exists \av \in \set{T}, \;\; \sum_{i=1}^t a_i e_i = 0   
		\right\} \\
		&\;\; \leq  \;\;
		\sum_{\av \in \set{T}} 
		\prob \left \{ \sum_{i=1}^t a_i e_i = 0   
		\right\}.
	\end{align*}
	Furthermore, because of \eqref{e_t=sum_iaiei}, the probability that  $\sum_{i=1}^t a_i e_i = 0 $ given that $(a_1,\dots,a_t) \in \set{T}$, is at most the ratio between the number of $(t-1)$-tuples that are linearly independent over the number of  linearly independent  $t$-tuples, that is
	\begin{align*}
		\prob \left \{ \sum_{i=1}^t a_i e_i = 0 \right\} 
		& \leq \dfrac{{\prod_{j=0}^{t-2} \left(q^m-q^j \right)}}{\prod_{j=0}^{t-1} \left(q^m-q^j \right)} = \frac{1}{q^{m} - q^{t-1}}.
	\end{align*}
The conclusion then follows as we have 
	\begin{align*}
		\sum_{\av \in \set{T}} 
		\prob \left \{ \sum_{i=1}^t a_i e_i = 0   
		\right\}
		&\;\; \leq  \;\;
		\left( q^{rt}- 1 - \left(q^{d} - 1\right) \sum_{u=0}^{t-1} q^u \right) \frac{1}{q^{m} - q^{t-1}}\\
		&\;\; \leq  \;\;
 		\frac{ q^{rt} - \left(q^d -1 \right) \frac{q^{t} - 1}{q-1} }{q^{m} - q^{t-1}} 
 		\;\; \leq  \;\;
 		\frac{ q^{rt} - \left(q^d -1 \right) \frac{q^{t} - 1}{q} }{q^{m} - q^{t-1}}
	\end{align*}
	which  provides the claimed bound \eqref{bound:aiei=0}.\qed
\end{proof}

\section{Decoding of LRPC Codes} \label{sec:decoding}

This section  is devoted to explaining how to  solve  efficiently the Rank decoding problem with an homogeneous matrix, or stated otherwise, we will explain how LRPC codes can be efficiently decoded. 
We first recall the definition of the \emph{Rank decoding} problem and then introduce the family of LRPC codes through the notion of homogeneous matrix.

\begin{definition}[Rank decoding problem]
	Let $q$, $m$, $n$, $k$, $t$  be a natural numbers such that $k <n$ and $t < n$.
	The \emph{Rank decoding} problem consists in finding $\ev$ from  the input $(  \Rm,  \ev  \Rm^{\mathsf{T}})$ assuming that $\Rm \in \MS{(n-k)}{n}{\fqm}$ and 
	$\ev \raffect \sphere_{t}(\fqm^n)$.
\end{definition}

\begin{definition}[Homogeneous matrix \& LRPC code]
	An  $r \times n$ matrix $\Mm$ is \emph{homogeneous of weight $w$ and support $\set{W}\in \grassman{w}{q}{m}$} if $\Mm \in \MS{r}{n}{\set{W}}$.
	A linear code  defined by an homogeneous parity-check matrix is named a \emph{Low Rank Parity Check (LRPC)} code. 
\end{definition}

Throughout this section we consider $\sv\in \fqm^{n-k}$, an homogeneous parity-check matrix $\Hm \in \MS{(n-k)}{n}{\fqm}$  of weight $w$ and support $\set{W}$ and an integer $t$. The goal is then to find  a vector $\ev  \in \fqm^n$ such that $\sv = \ev\Hm^\mathsf{T}$ and $\support{\fq}{\ev} \in \grassman{t}{q}{m}$.
Throughout this section we assume that an arbitrary basis $\{f_1,\dots,f_w\}$ of $\set{W}$ was picked, and the parameters
satisfy the following constraints, 
	\begin{equation} \label{eq:lrpc_constraints}
		\begin{cases}
			tw &\leq n- k, \\ 
			n  & \leq (n-k) w.
		\end{cases}
	\end{equation}

\subsection{Description}

We aim here to give a full description of the LRPC decoder. It consists of two steps that will be described below.   We will also give in Theorem~\ref{th:decoding}  
an upper-bound on the probability that the LRPC decoder fails. But before that, we first explain how from an input $(\Hm,\ev\Hm^\mathsf{T})$ the algorithm first recovers the support $\support{\fq}{\ev}$, and then all the entries of $\ev$.

\begin{algorithm}
	\begin{algorithmic}[1]
		\STATE{$\set{B} \affect \emptyset$}
		\IF{$\dim \bigcap\limits_{i = 1}^w  f_i^{-1}  \cdot \support{\fq}{\sv}  = t$}
			\STATE{$\set{B} \affect \{ \epsilon_1,\dots, \epsilon_t\}$ where $\epsilon_1,\dots, \epsilon_t$ is a basis of $\bigcap\limits_{i = 1}^w  f_i^{-1} \cdot \support{\fq}{\sv}$}
		\ENDIF  
		\RETURN{$\set{B}$}
	\end{algorithmic}
	\caption{\textsf{Step I -- Support Recovering $(\Hm,\sv,t)$}} \label{STEP1}
\end{algorithm}

The first step  is given in Algorithm \ref{STEP1}. The goal here is to compute a basis $\epsilon_1,\dots, \epsilon_t$ of  $\support{\fq}{\ev}$.
One can observe that the algorithm fails at this stage if one of the following events occurs:
\begin{enumerate}
		\item $\support{\fq}{\sv} \neq \set{E} \cdot \set{W}$ which will in particular always occur if $n - k < tw$,
		\item Or $\support{\fq}{\sv} = \set{E} \cdot \set{W}$ holds
		but yet the  strict inclusion  $\set{E} \; \subsetneq \; \bigcap_{i=1}^w  f_i^{-1} \cdot \support{\fq}{\sv}$ happens.
\end{enumerate}
In the following, we  elaborate more on these cases.
The second step then starts once a basis  $\epsilon_1,\dots, \epsilon_t$ of $\set{E}$ is  successfully recovered.
Next, it  checks whether the dimension of  $\set{E} \cdot \set{W}$ is equal to $t w$. Note that in this case, a basis of $\set{E} \cdot \set{W}$ is  given by
\[
\Big\{  f_i \epsilon_j  \;\; \big \vert\;\; i  \in \rg{1}{w},\;  j \in \rg{1}{t}   \Big\}.
\]
Each entry  of $\sv = \begin{bmatrix} s_{r} \end{bmatrix}$ is  written as 
$s_{r} = \sum_{i,j} \sigma^{(r)}_{i,j} f_i \epsilon_j$ where $\sigma^{(r)}_{i,j} $ lies in $\fq$.
Similarly each entry of $\Hm = \begin{bmatrix} h_{r,d} \end{bmatrix}$ with $d \in \rg{1}{n}$ is decomposed as 
$h_{r,d} = \sum_{i} \nu^{(r,d)}_{i} f_i$ with $\nu^{(r,d)}_{i} $ in $\fq$. Lastly each entry 
$e_{d}$ of the unknown vector $\ev$ is written as 
$e_{d} = \sum_{j} x^{(d)}_{j}   \epsilon_j$
where $x^{(d)}_{j}$  are unknowns that are sought in $\fq$ so  that we have
\begin{align*}
	s_{r} =\sum_{d=1}^n h_{r,d} e_{d} 
	&= \sum_{d=1}^n \left( \sum_{i=1}^w \nu^{(r,d)}_{i} f_i \right) \left( \sum_{j=1}^t x^{(d)}_{j}   \epsilon_j \right)\\
	&= \sum_{i=1}^w \sum_{j=1}^t \left(  \sum_{d=1}^n \nu^{(r,d)}_{i} x^{(d)}_{j} \right)  f_i \epsilon_j.
\end{align*}
The latter equality implies that we have a system of $(n-k) tw$ linear equations
involving $tn$ unknowns composed of the linear relations 
\[ 
\sigma^{(r)}_{i,j} = \sum_{d=1}^n \nu^{(r,d)}_{i} x^{(d)}_{j}
\]
where $(r,i,j)$ runs through $\rg{1}{n-k} \times \rg{1}{w} \times\rg{1}{t}$.
As we have taken $(n-k) w \geq n$ and since $\dim \set{E} \cdot \set{W} = tw$ we are sure to get a unique solution.   
We see in particular that this second step always fails if the dimension of $\set{E} \cdot \set{W}$ is not equal to $tw$.

\subsection{Decoding Failure Probability}

In this part, we focus on the question of estimating  the probability that the LRPC decoder fails on a random  input $(\Hm,\ev\Hm^\mathsf{T})$. Henceforth we denote it
by $\prob\Big \{\dec(\Hm, \ev\Hm^\mathsf{T}) \neq \ev \Big\}$ where $\dec$ denotes the LRPC decoder.
We also define the probability that $\dec$ fails at the first and second step by $\prob_{\rm I}$ and $\prob_{\rm II}$ respectively.
We then clearly have $\prob\Big \{\dec(\Hm, \ev\Hm^\mathsf{T}) \neq \ev \Big\} = \prob_{\rm I} + (1- \prob_{\rm I}) \prob_{\rm II}$ which implies that
\begin{align}
	\prob\Big \{\dec(\Hm, \ev\Hm^\mathsf{T}) \neq \ev \Big\}
	&\leq  \prob_{\rm I} +  \prob_{\rm II}. \label{Pfailure:leq:PI+PII}
\end{align}
We now state our main result.

\begin{theorem}\label{th:decoding}
	Consider natural numbers $w$, $t$, $m$, $k$, $n$ such that $tw\leq n-k$, $n \leq (n-k) w$ and $2(w - 1) t < m$.
	Assume that $\set{W} \raffect \grassman{w}{q}{m}$ and $\set{E} \raffect \grassman{t}{q}{m}$.
	For $\Hm \raffect \MS{(n-k)}{n}{\set{W}}$ and $\ev \raffect \set{E}^n$, the probability  $\prob\Big \{\dec(\Hm, \ev\Hm^\mathsf{T}) \neq \ev \Big\}$ is at most $\prob_{\rm I} +  \prob_{\rm II}$ where
	\[
	\begin{cases}
		\prob_{\rm I}	
		&\; \leq \; 
		1 -  \prod\limits_{i=0}^{tw-1} \left(1 - q^{i-(n-k)} \right)  + \dfrac{ q^{(2w-1)t+1} - \left(q^w -1 \right) \left(q^{t} - 1\right) }{q^{m+1} - q^{t}}, \\
		\prob_{\rm II}	
		&\;\leq \; \dfrac{q^{tw}}{q^m - q^{t-1}}.
	\end{cases}
	\]
\end{theorem}
The rest of this section is devoted to proving this theorem.

\subsection{An Upper-Bound on $\prob_{ \rm I}$}

The algorithm $\dec$ fails during the first step if either $ \support{\fq}{\sv} \neq \set{E} \cdot \set{W}$, or $\support{\fq}{\sv} = \set{E} \cdot \set{W}$ holds but we have $\set{E}  \neq  \bigcap_{i=1}^w  f_{i}^{-1} \cdot \support{\fq}{\sv}$.
Consequently the probability $\prob_{\rm I}$ is at most 
\begin{align*} 
\prob \Big \{   \support{\fq}{\sv} \neq \set{E} \cdot \set{W} \Big\}  
+ 
\prob \left \{ 
\set{E} 
\neq  
\bigcap_{i=1}^w  f_{i}^{-1} \cdot \support{\fq}{\sv}
\;\;  \Big \vert \;\; 
\support{\fq}{\sv} = \set{E} \cdot \set{W}
\right\}.
\end{align*}
In order to give an upper-bound on $\prob \Big \{\support{\fq}{\sv} \neq \set{E} \cdot \set{W} \Big\}$ we use Proposition  \ref{prop:randomUV} to claim that the entries of $\sv$ are independent and uniformly distributed random variables taking values on $\set{E} \cdot \set{W}$, and then we use Proposition  \ref{prop:probUnif} to bound the probability that randomly drawn vectors from a finite-dimensional vector space over $\fq$ form a set of maximum dimension.   

\begin{proposition} \label{prop:PI}
	For $\ev \raffect \set{E}^n$ and $\Hm \raffect \MS{(n-k)}{n}{\set{W}}$ where $\set{E} \in \grassman{t}{q}{m}$ and $\set{W} \in \grassman{w}{q}{m}$, the probability that $ \support{\fq}{\ev\Hm^\mathsf{T}}$ is different from 
	$ \set{E} \cdot \set{W}$ is
		\[
		\prob \left \{ \support{\fq}{\ev\Hm^\mathsf{T}}  \neq \set{E} \cdot \set{W} \right\} 
		\;\; \leq \;\; 
		1 - \prod_{i=0}^{tw-1} \left(1 - q^{i-(n-k)} \right). 
		\]
\end{proposition}

\begin{proof}
	Let $\hv_1,\dots,\hv_{n-k}$ be the rows of $\Hm$. 
	Consider a basis $\word{\epsilon}$ of $\set{E}$, and similarly fix an arbitrary basis $\word{\beta}$  of $\set{W}$.  
	Let us define $\Em \in \MS{t}{n}{\fq}$, and $\Mm_1 \in \MS{w}{n}{\fq},\dots,\Mm_{n-k} \in \MS{w}{n}{\fq}$ such that $\ev = \word{\epsilon} \Em$ and $\hv_i = \word{\beta} \Mm_i$ for each $i \in \rg{1}{n-k}$. Clearly the entries of the matrices $\Em$ and  $\Mm^\mathsf{T} \eqdef \begin{bmatrix} \Mm_1^\mathsf{T} &\vert & \cdots & \vert & \Mm_{n-k}^\mathsf{T}\end{bmatrix} \in \MS{n}{w(n-k)}{\fq}$ are independent and uniformly distributed random variables over $\fq$, and additionally, we have 
		\[
		\ev \Hm^\mathsf{T} =   \word{\epsilon} \Em \begin{bmatrix} \left(\word{\beta} \Mm_1\right)^\mathsf{T} & \cdots & \left(\word{\beta} \Mm_{n-k}\right)^\mathsf{T}\end{bmatrix} =  \word{\epsilon} \Em \Mm^\mathsf{T} 
		\begin{bmatrix} 
			\word{\beta}^\mathsf{T} & & \mat{0} \\
            & \ddots & \\ 
  			\mat{0} & & \word{\beta}^\mathsf{T}
		\end{bmatrix}.
		\]
	We know from Proposition \ref{prop:randomUV} that $\Mm \Em^\mathsf{T}$ is uniformly distributed matrix over $\MS{w(n-k)}{t}{\fq}$ which therefore implies that the entries of $ \ev \Hm^\mathsf{T}$ are independent and uniformly distributed random variables over $\set{E} \cdot \set{W}$. We then use Proposition \ref{prop:probUnif} to conclude.\qed
\end{proof}

We now focus on the second reason  why $\dec$ fails in step one, namely we would like to upper bound
$\prob \left \{ 
\set{E} 
\neq  
\bigcap_{i=1}^w  f_{i}^{-1} \cdot \support{\fq}{\sv}
\;\;  \Big \vert \;\; 
\support{\fq}{\sv} = \set{E} \cdot \set{W}
\right\} $.
This will be done in  Theorem~\ref{th:bound:intersection} whose proof requires Lemma \ref{lem:aiei=0} which will also be useful as we will see for establishing the probability of failure in the second step.

\begin{theorem} \label{th:bound:intersection}
	Let $\set{U} \eqdef \set{E} \cdot \set{W}$ where  $\set{W} \in  \grassman{w}{q}{m}$ and $\set{E} \raffect \grassman{t}{q}{m}$ with $(2w-1)t < m$.
	Then for an arbitrary basis $f_1,\dots{},f_w$ of $\set{W}$, we have
	\begin{equation*}
		\prob 
		\left \{ 
		\set{E} =   \bigcap_{i=1}^w  f_i^{-1}  \cdot \set{U} 
		\;\; 
		\Big \vert 
		\;\;
		\set{E} \raffect  \grassman{t}{q}{m}
		\right\} 
		\; 
		\geq
		\;
  		1 -   \frac{ q^{(2w-1)t+1} - \left(q^w -1 \right) \left(q^{t} - 1\right) }{q^{m+1} - q^{t}}\cdot
	\end{equation*}
\end{theorem}

\begin{proof}
	We know  that $\set{E} \neq \bigcap_{i=1}^w  f_i^{-1} \cdot \set{U}$ is actually equivalent to the strict inclusion $\set{E} \subsetneq   \bigcap_{i=1}^w f_i^{-1} \cdot \set{U}$, which in particular implies $\set{E} \subsetneq   f_1^{-1}  \cdot \set{U} \cap  f_2^{-1}  \cdot \set{U}$.
	Given a basis $e_1,\dots,e_t$ of $\set{E}$ and for every $j \in \rg{1}{w}$, a generating set of $f_j^{-1}  \cdot \set{U} $ as an $\fq$-linear subspace of $\fqm$ is given by  
	\begin{equation}
		\Big \{ e_1,\dots,e_t \Big \} 
		\; \bigcup \; 
		\Big \{
		e_k f_\ell  f_j^{-1}  \;\; \Big\vert \;\; \ell \in \rg{1}{w} \setminus \{j \}, \; k \in \rg{1}{t} 
		\Big\}.
	\end{equation}
	So the existence of a non-zero element in
	$ f_1^{-1}  \cdot \set{U} \cap  f_2^{-1} \cdot \set{U}$ means that there exist scalars  $\lambda_k$, $\gamma_k$, $\alpha_{k,\ell}$, $\beta_{k,j}$  in $\fq$ not all zero such that
	\begin{equation} \label{eq:intersection:f1f2}
		\sum_{k=1}^t \lambda_k e_k + \sum_{k=1}^t \sum_{\ell= 2}^w  \alpha_{k,\ell} f_1^{-1}f_\ell e_k = \sum_{k=1}^t \gamma_k e_k + \sum_{k=1}^t \sum_{j =1 , j\neq 2}^w \beta_{k,j} f_2^{-1}f_j e_k.
	\end{equation}
	In order to have this element not in $\set{E}$, we must have in particular $\alpha_{k,\ell}$ and $\beta_{k,j}$ not all zero.
	In other words, by defining  $\set{A}$ as the subset of $\fqm$ such that 
	\[
	\set{A} \eqdef \left\{\lambda  + \sum_{\ell = 2}^w  \alpha_{\ell} f_1^{-1}f_\ell 
	+ \sum_{j =1, j \neq 2}^w \beta_{j} f_2^{-1}f_j \;\; \Big\vert \;\; \lambda\in\fq,\;  
	(\alpha_{\ell},\beta_{j})\in\fq^{2w-2}\setminus \{\word{0}\} \right\}
	\] 
	we see that \eqref{eq:intersection:f1f2} entails that there exist $a_1,\dots,a_t$ in $\set{A}$ such that  $\sum_{k=1}^t e_k a_k = 0$.
	Notice also that $\set{A}$ is included inside a linear space of dimension $2w-1$ and contains the linear space of dimension $w$ that is generated by the linearly independent elements $1,f_1^{-1}f_2,\dots,f_1^{-1}f_w$.
	Consequently  from Lemma~\ref{lem:aiei=0} we can write that
	\begin{align*}
		\prob 
		\left \{ 
		\set{E} \neq   \bigcap_{i=1}^w  f_i^{-1}  \cdot \set{U} 
		\;\; 
		\Big \vert 
		\;\;
		\set{E} \raffect  \grassman{t}{q}{m}
		\right\} 
		\; 
		&\;\; \leq \;\;
		\prob 
		\Big \{ 
		\set{E} \subsetneq   f_1^{-1}  \cdot \set{U} \cap  f_2^{-1}  \cdot \set{U} 
		\Big\}  \\
		&
		\;\; 
		\leq
		\;\;
		\frac{ q^{(2w-1)t+1} - \left(q^w -1 \right) \left(q^{t} - 1\right) }{q^{m+1} - q^{t}}
	\end{align*}
	which concludes the proof.
	\qed
\end{proof}

\subsection{An Upper-Bound on $\prob_{\rm II}$}

We have seen that  the second step of $\dec$ fails if the dimension of $\set{E} \cdot \set{W}$ is not equal to $t w$, that is to say we have
\begin{equation} 
	\prob_{\rm II}
	= \prob\Big\{ \dim \; \set{E} \cdot \set{W} \ne tw \Big\}.
\end{equation}
Then the bound given in Theorem \ref{th:decoding} follows from the following result that can be proved thanks to Lemma~\ref{lem:aiei=0}.

\begin{proposition} \label{prop:dimEW = wt}
	For   $\set{W} \in \grassman{w}{q}{m}$ and assuming that $wt < m$, we have 
	\begin{equation} \label{eq:dimEW=wt}
		\prob\Big\{\dim \set{E} \cdot \set{W}  = tw \;\; \Big \vert \;\; \set{E} \raffect  \grassman{t}{q}{m}\Big\} 
		\; \geq \;   
		1 -  \frac{q^{tw}}{q^m - q^{t-1}}.
	\end{equation}
\end{proposition}

\begin{proof}
	$ \set{E} \cdot \set{W} $ is generated by $ \{e_i f_j \; \vert \; 1 \leq i \leq t, \;  1 \leq j \leq w \}$ where $\{f_j \; \vert \; 1 \leq j \leq w \}$ is a basis for $\set{W} $ and $\{e_i \; \vert \; 1 \leq i \leq t \}$ is a basis for $\set{E} $. Furthermore, the dimension of $\set{E} \cdot \set{W}$ is different from $tw$ means that there exist scalars $\gamma_{i,j}$ in $\fq$ such that
	\[
	\sum_{i = 1}^t \left( \sum_{j=1}^w \gamma_{i,j} f_j \right) e_i =0.
	\] 
	We can then apply Lemma~\ref{lem:aiei=0} in order to obtain the following lower-bound 
	\[
	\prob\Big\{\dim \set{E} \cdot \set{W}  \neq tw \;\; \Big \vert \;\; \set{E} \raffect  \grassman{t}{q}{m}\Big\} 
	\; \leq \;   
	\frac{q^{tw}}{q^m - q^{t-1}}.
	\]
	This clearly is equivalent to  \eqref{eq:dimEW=wt} and terminates the proof.
	\qed
\end{proof}

\section{Asymptotic Analysis}

We recall from Proposition \ref{prop:PI} that the probability that the coordinates of $\ev\Hm^\mathsf{T}$ do not span $\set{E} \cdot \set{W}  $ is given
\begin{equation}\label{eq:probPI2}
	\prob \left \{ \support{\fq}{\ev\Hm^\mathsf{T}}  \neq \set{E} \cdot \set{W} \right\} 
	\;\;
	\leq
	\;\;
	 1 - \prod_{i=0}^{tw-1} \left(1 - q^{i-(n-k)} \right). 
\end{equation}
The goal here is to upper-bound the term $ 1 - \prod_{i=0}^{tw-1} \left(1 - q^{i-(n-k)} \right)$. 

\begin{proposition} \label{prop:simplified_prod_bound} 
	Let us define  $T(q,t,w) \eqdef 1 - \prod_{i=0}^{tw-1} \left(1 - q^{i-(n-k)} \right)$.
	Then under the condition that $twq^{-(n-k)+tw}\leq 1$, we have
		\[
		0	\;\; \leq \;\; 
		\frac{q^{-(n-k)+tw}}{q-1}  - T(q,t,w)
		\;\; \leq \;\; 
		\frac{q^{-(n-k)}}{q-1} + \frac{1}{q+1}\left(\frac{q^{-(n-k)+tw}}{q-1}\right)^2
		\]
	In particular, with $tw = \omega(1)$ and $k = \Theta(n)$, it entails that $T(q,t,w) \sim \frac{q^{-(n-k)+tw}}{q-1}$ as $n$ tends to $+\infty$.
\end{proposition}

\begin{remark}\label{rk:equiv}
	We know by Remark \ref{rk:a=d} that if $\dim(\set{E} \cdot \set{W}) = tw$, then \eqref{eq:probPI2} is an equality. Hence $\prob \left \{ \support{\fq}{\ev\Hm^\mathsf{T}}  \neq \set{E} \cdot \set{W} \right\}$ is equivalent to $q^{-(n-k)+tw}/(q-1)$.
\end{remark}
	
\begin{proof}
	Note that by expanding the expression of $1 - \prod_{i=0}^{tw-1} \left(1 - q^{i-(n-k)} \right)$ we have
	\begin{equation} \label{Tq:sum_ui}
		T(q,t,w) = 1 - \prod_{i=0}^{tw-1} \left(1 - q^{i-(n-k)} \right) = \sum_{i=1}^{tw} (-1)^{i+1} u_i
	\end{equation}
	where
	\[
	u_i \eqdef \sum_{{\{k_1,\dots,k_i\} \subseteq \rg{0}{tw-1}\atop \card{\{k_1,\dots,k_i\}} = i}} q^{\sum_{j=1}^{i} k_j - i(n-k)}.
	\]
	We will prove that the sequence $(u_i)_{0\leq i \leq tw - 1}$ is decreasing. Let us suppose for the moment that it is true. 
	Then the whole sum  $\sum_{i=1}^{tw} (-1)^{i+1} u_i$   satisfies the inequalities
	\begin{equation} \label{eq:u1_u2}
		u_1 - u_2
		\;\; \leq \;\;
		\sum_{i=1}^{tw} (-1)^{i+1} u_i 
		\;\; \leq \;\;
		u_1
	\end{equation}
	with  $u_1 =  q^{-(n-k)} \sum_{j=0}^{tw-1}q^j = q^{-(n-k)}(q^{tw}-1)/(q-1)$  and $u_2 \leq \frac{q^{-2(n-k)+2tw}}{(q-1)^2}$ because of the following series of inequalities,			
	\begin{align*}
		u_2 
		= q^{-2(n-k)} \sum_{k_1 = 0}^{tw-2}q^{k_1}\sum_{k_2 = k_1+1}^{tw-1}q^{k_2} 
		&=\frac{q^{-2(n-k)}}{q-1} \sum_{k_1 = 0}^{tw-2}q^{k_1}  \left(q^{tw} - q^{k_1+1}\right) \\
		&= \frac{q^{-2(n-k)}}{(q-1)^2} \left( q^{2tw-1} - q^{tw} -  \frac{q^{2tw-1 } - q }{q+1} \right)\\
		&= \frac{q^{-2(n-k)}}{(q+1)(q-1)^2} \left( q^{2tw} - (q  + 1) q^{tw} + q  \right)\\
		&\leq \frac{ q^{-2(n-k)+2tw} }{(q+1)(q-1)^2} 
	\end{align*}
	Gathering this last inequality with \eqref{Tq:sum_ui} and \eqref{eq:u1_u2}, we obtain then
		\[
 		\frac{q^{-(n-k)}}{q-1}\left(q^{tw}-1\right)
		-
		\frac{q^{-2(n-k)+2tw}}{(q+1)(q-1)^2}\;\; \leq \;\; T(q,t,w)  \; \; \leq \;\; \frac{q^{-(n-k)}}{q-1}\left(q^{tw}-1\right)
		\]
	To finish the proof, it only remains to prove that  $u_1,\dots,u_{tw}$ is a decreasing sequence. Let us choose $i$ in $\rg{1}{tw-1}$. We have then
	\begin{align*}
		u_{i+1} &= q^{-(i+1)(n-k)}\sum_{{\{k_1,\dots,k_{i+1}\} \subseteq \rg{0}{tw-1}\atop \card{\{k_1,\dots,k_{i+1}\}} = i+1}} q^{\sum_{j=1}^{i+1} k_j}\\
		&= q^{-i(n-k)} q^{-(n-k)+tw} 
		\sum_{{\{k_1,\dots,k_{i+1}\} \subseteq \rg{0}{tw-1}\atop \card{\{k_1,\dots,k_{i+1}\}} = i+1}} 
		q^{k_{i+1} - tw + \sum_{j=1}^{i} k_j }\\
		&\leq q^{-(n-k)+tw} q^{-i(n-k)} \sum_{{\{k_1,\dots,k_{i+1}\} \subseteq \rg{0}{tw-1}\atop \card{\{k_1,\dots,k_{i+1}\}} = i+1}} q^{\sum_{j=1}^{i} k_j}\\
		&\leq q^{-(n-k)+tw} q^{-i(n-k)} \sum_{\ell =0}^{tw-1} \sum_{{\{k_1,\dots,k_{i}\} \subseteq \rg{0}{tw-1}\setminus\{\ell\}\atop \card{\{k_1,\dots,k_{i}\}} = i}} q^{\sum_{j=1}^{i} k_j}\\
		&\leq twq^{-(n-k)+tw} u_i.
	\end{align*}
	By assumption we have $ twq^{-(n-k)+tw} \leq 1$ which shows that $u_{i+1} \leq u_i$. 
	\qed	
\end{proof}

\begin{corollary}\label{cor:asymp}
	With $twq^{-(n-k)+tw}\leq 1$, $tw = \omega(1)$ and $k = \Theta(n)$, we have when $n\rightarrow\infty$
	\[
	\prob\Big \{\dec(\Hm, \ev\Hm^\mathsf{T}) \neq \ev   \Big\}\; \leq\; \dfrac{q^{-(n-k)+ tw}}{q-1}+ q^{2tw -m}
	\]
\end{corollary}

\begin{proof}
	It follows from Theorem \ref{th:decoding} and Proposition \ref{prop:simplified_prod_bound}.
\end{proof}

\section{Conclusion}

The LRPC decoding algorithm is becoming  more and more  a predominant  tool in rank-metric cryptography as it is the main ingredient that serves to invert encryption functions in \cite{ABDGHRTZABBBO19,BGHO22,AADGZ22}. It is therefore of great  importance to establish well-grounded  bounds on the decoding failure probability to ensure a trust on the parameters provided for those schemes. Yet all existing bounds are either too loose for being interesting in concrete cryptographic applications, or are tight according to  experimental observations but  are not supported by realistic model. 
This work partially fill this gap by improving existing theoretical bounds. Our upper-bound behaves asymptotically as $q^{-(n-k)+ tw}/(q-1)+ q^{2tw -m}$.

However, there is still a large gap with the experimental bound given in \cite{ABDGHRTZABBBO19,AADGZ22} that comes from the second case of failure in the first step of the decoding algorithm. 
That is why a finer analysis of this event could result to a better bound.
Lastly, our analysis applies specifically to ``unstructured'' LRPC codes and it would be interesting to study the decoding failure probability of \emph{ideal} LRPC codes.

\bibliographystyle{plain}

\end{document}